\documentclass[showpacs,showkeys,preprintnumbers,fleqn]{revtex4}
\usepackage{amsmath,amssymb}
\usepackage{graphicx}
\usepackage{bm}

\def\bra{\langle}
\def\ket{\rangle}
\def\qbar{\overline{{\rm q}}}

\def\ubar{\overline{{\rm u}}}
\def\dbar{\overline{{\rm d}}}
\def\sbar{\overline{{\rm s}}}
\def\Kbar{\overline{{\rm K}}}
\def\rmu{{{\rm u}}}
\def\rmd{{{\rm d}}}
\def\rms{{{\rm s}}}

\def\rmN{{{\rm N}}}

\def\toprule{\hline}

\def\Voge{{V_{\rm OGE}}}

\def\Vconf{{V_{\rm conf}}}

\def\aconf{{a_{\rm conf}}}

\def \mbf#1{\mbox{\boldmath{$#1$}}}
\def \vecp{\mbf{p}}

\def \vecr{\mbf{r}}

\def\lamilamj{(\lambda_i\cdot\lambda_j)}
\def\sigisigj{(\mbf{\sigma}_i\cdot\mbf{\sigma}_j)}

\newcommand{\FRAC}[2]{\leavevmode\kern.1em
  \raise.5ex\hbox{\the\scriptfont0 #1}\kern-.1em
  /\kern-.15em\lower.25ex\hbox{\the\scriptfont0 #2}}
\newcommand{\xbld}[1]{\mbox{\boldmath $ #1 $}}

\def\Voge{V_{\rm OGE}}

\def\Viii{V_{\rm INS}}

\def\Viiit{V_{\rm INS}^{(2)}}
\def\Viiitv{V_{\rm INS}^{(2a)}}
\def\Viiih{V_{\rm INS}^{(3)}}
\def\Viiihv{V_{\rm INS}^{(3a)}}
\def\aiiit{V_{0}^{(2)}}
\def\aiiih{V_{0}^{(3)}}
\def\Uiiit{U_{ij}^{(2)}}
\def\Uiiih{U_{ijk}^{(3)}}
\def\Uiiihv{U_{ij5}^{(3)}}

\def\Vconf{V_{\rm Conf}}
\def\aconf{a_{\rm Conf}}
\def\veck{\bm{k}}
\def\Pins{{\cal P}_{\rm INS}}

\def\tlambda{\tilde{\lambda}}
\def\tsigma{\tilde{\sigma}}

\def\lamlam{\lambda_i\!\cdot\!\lambda_j}
\def\sigsig{\sigma_i\!\cdot\!\sigma_j}

\def\FF{($F\!\cdot\! F$)}

\def\lamjlamk{\lambda_j\!\cdot\!\lambda_k}
\def\lamklami{\lambda_k\!\cdot\!\lambda_i}
\def\sigjsigk{\sigma_j\!\cdot\!\sigma_k}
\def\sigksigi{\sigma_k\!\cdot\!\sigma_i}
\def\tlamlam{\tlambda_i\!\cdot\!\tlambda_j}
\def\tlamjlamk{\tlambda_j\!\cdot\!\tlambda_k}
\def\tlamklami{\tlambda_k\!\cdot\!\tlambda_i}
\def\tsigsig{\tsigma_i\!\cdot\!\tsigma_j}
\def\tsigjsigk{\tsigma_j\!\cdot\!\tsigma_k}
\def\tsigksigi{\tsigma_k\!\cdot\!\tsigma_i}

\def\E0pole{E^{(0)}_{pole}}
\def\half{{1\over 2}}

\begin{document}
\setlength{\baselineskip}{15pt}

\title{$\Lambda$(1405) as a Resonance in the Baryon-Meson Scattering Coupled to the q$^3$ State \\
in a Quark Model}
\author{Sachiko Takeuchi}\affiliation{%
Japan College of Social Work, Kiyose, Tokyo 204-8555, Japan}
\author{Kiyotaka Shimizu}\affiliation{%
Department of Physics, Sophia University, Chiyoda-ku, Tokyo 102-8554, Japan
}
\date{\today}


\begin{abstract}
In order to describe $\Lambda$(1405) as a resonance in the
baryon-meson scattering, we have investigated  q$^3$-q$\qbar$ scattering
system with the flavor-singlet q$^3$ $(0s)^2(0p)$ state (the $\Lambda^1$ pole).
The scattering is treated by the 
quark cluster model (QCM).
The $\Lambda^1$ pole
is treated as a bound state embedded in the continuum.
We found that the peak appears below the N$\overline{\rm K}$ threshold
in the spin ${1\over 2}$, isospin 0 channel even if the mass of the $\Lambda^1$ pole
is above the threshold.
This peak disappears when the coupling to the $\Lambda^1$ pole is switched off.
To use the observed hadron mass in the kinetic part of QCM is also found
to be important to reproduce a peak just below the N$\Kbar$ threshold.
\end{abstract}

\maketitle

\section{Introduction}

Constituent quark models are known to reproduce the observed low-energy features of 
baryons.
The number of ground state baryons corresponds to the observed one.
The masses and magnetic moments of the flavor-octet and decuplet baryons
are also reproduced well.
The picture that the baryon mass comes mainly from the masses of the 
three constituent
quarks,
which interact with each other by exchanging gluons and/or Goldstone bosons,
seems appropriate\cite{Isgur:1978xj,Isgur:1992td,Karl:1991kp,gloz,FS02,FST03}. 

There, however, are some exceptions. One of these is that such models cannot give the observed light mass of
$\Lambda(1405)$ nor the large mass difference between 
$\Lambda(1405,\frac{1}{2}^-)$ and $\Lambda(1520,\frac{3}{2}^-)$.
In the conventional constituent quark model, each of 
these baryons is treated as 
a system of three quarks in the flavor-singlet $(0s)^2(0p)$ state \cite{Isgur:1978xj}.
Then,
the hyperfine interaction only gives a part ($\sim$ 150 MeV) 
of the observed mass difference between 
the flavor-singlet and the flavor-octet spin-${1\over 2}$ baryons, $\sim$ 200 MeV.
Moreover, in order 
to give the large mass difference between 
$\Lambda(1405)$ and $\Lambda(1520)$,
one has to assume a strong spin-orbit force
between quarks, which is absent in other negative-parity baryons.

The excited baryons are embedded in the baryon-meson scattering states.
Therefore it is most appropriate to investigate $\Lambda(1405)$ as a meson-baryon resonance. 
%
In ref.\ \cite{Yamazaki:2002uh}, the idea that $\Lambda$(1405) is a N$\Kbar$ bound state
has been presented. 
A more systematic treatment may be found in
the chiral unitary approach, where
the lowest order baryon-meson vertices of the nonlinear chiral Lagrangian
are employed as the baryon-meson 
interaction \cite{Oset:1997it,Hyodo:2003jw,Magas:2005vu,Jido:2003cb,Jido:2005ew}. 
Because the model Lagrangian is an flavor-SU(3) extended version, 
this interaction is flavor-flavor type; the relative strength of 
the potentials among the  
 SU(3) octet baryons and mesons is governed by the SU(3) relation. 
The interaction is very short-ranged and 
 strongly attractive in both of the $\Sigma\pi$ and N$\Kbar$
$(TJ)=(0{1\over 2})$ channels.
 Furthermore, it is easily seen that this flavor-flavor type interaction 
 gives the strongest attraction for the flavor-singlet state among the
baryon-meson system. This is essentially the origin to 
produce the $\Lambda$(1405) resonance in this scheme.

There are also some attempts to express this baryon-meson picture by using quark models.
To treat the negative-parity baryons as q$^4\qbar$ systems
was proposed in ref.\cite{Hogaasen:1978jw}. 
This idea, however, has been abandoned because it was found 
by the  q$^4\qbar$ $(0s)^5$ calculations that 
some of the other flavor-octet negative-parity baryons are also found to have light mass.
There is a report to examine the N$\Kbar$ scattering from a quark model 
\cite{Wang:2004ky}.
Recent progress in the calculation method for few-body systems allows
us to solve a q$^4\qbar$ system by taking the orbital correlation fully into account.
Then, it is found that the one-gluon exchange potential (OGE) \cite{DeRujula:1975ge}, 
which is strongly attractive 
in the $\Sigma\pi$ channel,
causes a bound state below the $\Sigma\pi$ threshold in this $(TJ)=(0{1\over 2})$ channel
 \cite{Nakamoto:2006br}.
Since there is no strong attraction from color-magnetic interaction (CMI)
 in the N$\overline{\rm K}$ channel,
it seems unlikely to have a resonance close to but below the N$\Kbar$ threshold in this approach.
To explain $\Lambda$(1405) by the baryon-meson picture directly using the quark models 
does not seem to work so far.

An example to combine the above two pictures, viz., 
the q$^3$ picture and the baryon-meson one, 
is found in ref. \cite{arima}, where the meson-cloud effects   
on the flavor-singlet q$^3$ baryons were investigated.
It was reported that the self-energy of the 
q$^3$ state makes the baryon mass considerably lower.
It seems necessary to introduce such a q$^3$ state, in addition to the baryon-meson state,
for the constituent quark models to 
express the negative-parity baryons.


In this paper, in order to clarify the above situation 
of $\Lambda$(1405),
we perform the dynamical calculations of the 
q$^3$-q$\qbar$ scattering systems 
with a flavor-singlet q$^3\ (0s)^20p$ state as an embedded pole in the continuum, 
which we call `the $\Lambda^1$ pole' in the following.
For this purpose, 
we employ the quark cluster model (QCM),
which has succeeded in describing the 
baryon-baryon scattering quantitatively,
and study the baryon-meson scattering in this paper \cite{okya,Shimizu:1989ye,supl}. 
There are two points which are taken into account especially in this
baryon-meson scattering problem. 
One of them, as was mentioned above, is a coupling to the $\Lambda^1$ pole
in addition to the usual coupled-channel baryon-meson scattering calculation. 
The other one is
a modification of the relative kinetic energy 
of QCM. 
In the nonrelativistic scheme of the quark models, the reduced mass of the baryon-meson system 
is 
$\mu=6m_q/5$ with the constituent quark mass, $m_q$. 
When we study the $\Lambda$(1405) 
in terms of the baryon-meson scattering, 
the most important channel is $\Sigma \pi$ channel.
In reality, it has much smaller reduced mass, 124 MeV, compared with that of QCM, 
which is roughly 400 MeV. 
The kinetic energy is underestimated severely.
To remove this, 
we replace the reduced mass in QCM by the real one;
we have found that to take this effects into account is essential to discuss the baryon-meson scatterings
by the quark model.

%
As for the interaction between quarks, we follow the most common approach based on OGE and 
linear confinement. This model nicely reproduces the SU(3) octet and decuplet baryons.
In order to describe the SU(3) octet and singlet meson spectra, we include the instanton 
induced interaction (III) 
\cite{'tHooft:1976fv,Shifman:1979uw,Shifman:1979nz,
Shuryak:1984nq,Shuryak:1989cx,Kochelev:1985de,Callan:1976je,
Shuryak:1988bf,Oka:1990vx,Oka:1989ud,Takeuchi:1990qj,Takeuchi:1994ma,Morimatsu:1992mm,Takeuchi:1993rs}.
We also introduce a purely phenomenological potential to lower the 
pseudoscalar SU(3) octet mesons, which is considered to express the collective nature of the 
Goldstone bosons.
Employing this model, we can reproduce the observed baryon and
 meson masses well after an assumption
for the zero-point energy is introduced.
We use this `OGE model' rather than the `chiral model,' in which the quarks interact with each other
by exchanging the Goldstone bosons.
The OGE model may not produce the long-range behavior of the system with a lack of meson-exchange
between the baryon and meson.  
We, however, use this model as a first step, 
because it is very difficult for the chiral model to reproduce the meson spectrum,
and 
because it is not clear how to remove a possible overcounting 
from the treatment which handles the scattering q$\qbar$ mesons 
and the exchanged mesons simultaneously if we use the chiral model for the baryon-meson scattering problem.

In the next section, we explain the model hamiltonian. 
Then in section 3, a brief summary of QCM for baryon-meson scattering will 
 be given. 
There we explain how to modify the relative kinetic energy between baryon and meson in the quark 
 cluster model, and how to take into account the $\Lambda^1$ pole in QCM. 
 Numerical results for the coupled channel $\Sigma \pi$ and N$\Kbar$ scattering  and discussion are given 
 in section 4. Summary is given in section 5.   

\section{Hamiltonian}

We have employed a valence quark model. 
 The hamiltonian is taken as:
\begin{equation}
H_q = \sum_i (\frac{p_i^2}{2 m_i}  +V_0) -K_G
+\sum_{i<j} \left(\Voge{}_{ij} + \Vconf{}_{ij}\right) 
+ \sum_{i=1}^4 (V^{(a)}_{{\rm OGE}i5} + 
V_{{\rm coll}i5}) +\Viii.
\end{equation}
Here, $K_G$ is the kinetic energy from the center of mass motion.
We take the zero-point energy, which is (the number of quarks) $\times V_0$.

The two-body potential term consists of the one-gluon-exchange potential, 
$\Voge$ and $\Vconf$, which are defined as:
%
\begin{eqnarray}
\Voge{}_{ij} &=&\lamilamj{\alpha_s\over 4}\left\{
{1\over r_{ij}}
\!-\! \left({\pi\over 2 m_i^2}\!+\!{\pi\over 2 m_j^2}\!+\!{2\pi\over 3 m_im_j}\sigisigj \!\right)
\delta (\xbld{r}_{ij})
\right\},
\\ 
{\Vconf}_{ij} &=& 
-\lamilamj \; \aconf \;r_{ij},
\label{eq:conflam} 
\end{eqnarray}
where $\lambda_i$, $\mbf{\sigma}_i$, $m_i$, and $\xbld{r}_{ij}$ are color SU(3) generator, 
Pauli spin operator, the mass for the $i$-th quark, and the 
distance between the $i$-th and $j$-th quarks, respectively. 
The $\alpha_s$ and $\aconf$ are 
the strong coupling constant and the confinement strength. 
%
$V^{(a)}_{{\rm OGE}i5}$ is the term due to the pair annihilation, and $V_{\rm coll}$ is 
introduced to 
act only on the color-singlet, pseudoscalar SU(3) octet mesons:
\begin{eqnarray}
V^{(a)}_{{\rm OGE}i5}&=&
{1\over 24}\Big({16\over 3}-\lambda_i 
\!\cdot\!\lambda_{5}^{\ast}\Big)
\big(3+\xbld{\sigma}_i \cdot \xbld{\sigma}_5 \big)
{\cal P}_{i5}
\;{\pi \alpha_s} {1\over 4 m_{u}^2}\;
\delta (\xbld{r}_{i5})
\\
V_{{\rm coll}i5}&=&\frac{v_0}{24}(1-\xbld{\sigma}_i \cdot \xbld{\sigma}_5)
(\frac{16}{3}-f_i \cdot f_5^{\ast})
(\frac{1}{9}+\frac{\lambda_i \cdot \lambda_5^{\ast}}{6})\delta (\xbld{r}_{ij}),
\end{eqnarray}
where $f_i$ is the flavor SU(3) generator for the $i$-th quark, and 
the antiquark is denoted as the 5-th quark. The flavor operator ${\cal P}_{i5}$ has a matrix 
element among u$\ubar$, d$\dbar$ and s$\sbar$ for the $i$-th and 5-th quarks. 
The explicit form of the matrix elements is as follows.
\[
\left ( \begin{array}{ccc}
1 & 1 & \xi_s \\
1 & 1 & \xi_s \\
\xi_s & \xi_s & \xi_s^2 
\end{array}
\right )  \hspace{10pt}\mbox{with}~~~ \xi_s=\frac{m_u}{m_s},
\]
where $m_u$ is the mass of u and d quarks and $m_s$ is the mass of s quark. 
Note that the operator ${\cal P}_{i5}$ has a non-zero matrix element only
between a flavor singlet q$\qbar$ state,
 namely, $(\rmu \ubar + \rmd \dbar +\rms \sbar)/\sqrt{3}$ state when $\xi_s=1$. 
$V_{{\rm coll}}$, which acts on a spin-0 flavor-octet color-singlet q$\qbar$ pair, stands for the collective nature of 
Goldstone bosons. The form of this term is determined by purely phenomenologically.

$\Viii$ is the instanton-induced interaction which stands for the
short-ranged nonperturbative effects of the gluon field
 \cite{Oka:1990vx,Oka:1989ud,Takeuchi:1990qj,Takeuchi:1994ma,Morimatsu:1992mm,Takeuchi:1993rs}: 
\begin{eqnarray}
\Viii&=&\Viiit+\Viiitv+\Viiih+\Viiihv.
\end{eqnarray}
Here, $\Viiit+\Viiitv$ and $\Viiih+\Viiihv$ are the two- and three-body parts of $\Viii$, respectively.
$\Viiitv$ corresponds to the two-body term due to the pair annihilation,
which gives the $\eta$-$\eta'$ mass difference.
$\Viiihv$ corresponds to the three-body term also due to the pair annihilation.
Their explicit forms are
\begin{eqnarray}
\Viiit&=&{\aiiit \over 2} \sum_{i<j}
\xi_i\xi_j\Pins \,\Uiiit \,\delta^3(\vecr_{ij})\label{eq:viiiqq}
\\
\Viiitv &=&-{\aiiit \over 2}\sum_{i} \Pins {\cal P}_{i5}\,\Uiiit\,\delta^3(\vecr_{i5})
\\
\Viiih&=&{\aiiih \over 4} \sum_{i<j<k}
\Pins\, \Uiiih\,
\delta^3(\vecr_{ij})\delta^3(\vecr_{ik})\label{eq:viiiqq}
\\
\Viiihv &=&-{\aiiih \over 4}\sum_{i<j}\Pins\,\Uiiihv\,
 \delta^3(\vecr_{ij})\delta^3(\vecr_{i5})
\end{eqnarray}
with 
$\xi_i=m_u/m_i$.
$\aiiit$ [$\aiiih$]  is the effective strength 
of the two- [three-] body part of the $\Viii$ potential, which are related each other by the
quark condensate:
\begin{eqnarray}
\aiiit = \bra \ubar {\rm u} \ket {1\over 2 \xi_s} \aiiih.
\end{eqnarray}
Due to the projection operator in the flavor space, $\Pins$, these terms
do not vanish only if all the incoming quark lines
have different flavor from each other: it allows
 u$\ubar \rightarrow {\rm d}\dbar, {\rm s}\sbar$, {\it etc.}\
for the two-body part, and ud$\dbar \rightarrow {\rm us}\sbar$,  {\it etc.}\ for the three-body part,
but forbids u$\ubar \rightarrow {\rm u}\ubar$ or uu $\rightarrow$ uu and so on.

The spin-color part of the two-body interaction is:
\begin{eqnarray}
\Uiiit &=& 1+{3\over 32}\tlamlam
+{9\over 32}\lamlam \sigsig \label{eq:uiiiqq}
\end{eqnarray}
with $\tlambda=\lambda$ for a quark and $\lambda^*$ for an antiquark.
Also with $\tsigma=\sigma$ for a quark and ($-\sigma$) for an antiquark, 
the three-body part is expressed as 
\begin{eqnarray}
\Uiiih&=&1+{3\over 32}(\tlamlam
+\tlamjlamk+\tlamklami)
+{9\over 320}d_{abc}\tlambda_i^a\tlambda_j^b\tlambda_k^c
\nonumber \\
&+&{9\over 32}(\lamlam \sigsig+\lamjlamk \sigjsigk+\lamklami \sigksigi) 
+{27\over 320}d_{abc}\tlambda_i^a\tlambda_j^b\tlambda_k^c
(\tsigsig+\tsigjsigk+\tsigksigi)
\nonumber \\
&-&{9\over 64}f_{abc}\epsilon_{\ell mn}\lambda_i^a\lambda_j^b\lambda_k^c
\sigma_i^\ell\sigma_j^m\sigma_k^n
\Big), \label{eq:uiiiqqq}
\end{eqnarray}
where $\epsilon_{\ell mn}$ is the structure constant of SU(2) group, and $f_{abc}$ and
$d_{abc}$ are the structure constants of SU(3) group.

\begin{figure}[tp]
\caption{\label{fig:diag1}Annihilation diagram of the
$j$-th quark and the $\overline{k}$-th antiquark by OGE,
where the gluon carries three-momentum $\veck=\vecp_j+\vecp_{\overline{k}}$ to the $i$-th quark.}
\includegraphics[scale=0.5]{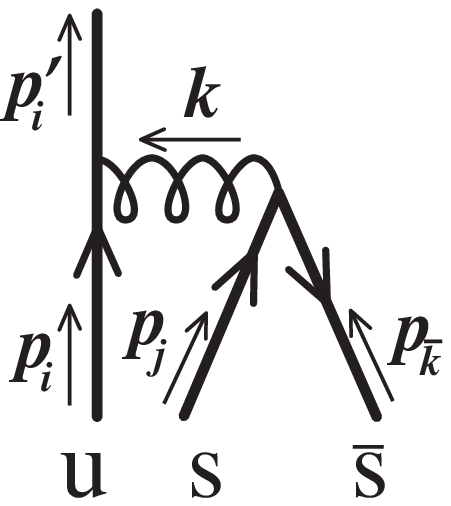}
\end{figure}
We assume that the coupling between the scattering state and the $\Lambda^1$ pole comes from OGE. (See Fig.\ \ref{fig:diag1}.)
Since the size of $\Viii$ is empirically found to be comparable to CMI,
which is the next order term, we only consider the leading term of OGE for this coupling.
The term can be written as:
\begin{eqnarray}
\bra V_{3q-5q}\ket &=&6\;\bra q^3|V_{3;4\overline{5}}{1\over\sqrt{4}}(1-2P_{24}-P_{34})|B(q^3)_{123}M(q_4\qbar_5)\ket 
\end{eqnarray}
with
\begin{eqnarray}
V_{i;j\overline{k}} &=& \lambda_i\cdot \lambda_{j\overline{k}}\;
{\alpha_s\over 4} {\pi\over m_a^2} \Big\{\Big({1\over 2}\big(
{1\over  m_a}+{1\over  m_i}\big)\veck
-{i \sigma_i\times\veck \over 2m_i}\Big)\cdot \sigma_{j\overline{k}}
\Big\},
\label{eq:V3q5q}
\end{eqnarray}
where $m_a$ is the quark mass in the annihilating q$_j\qbar{}_{\overline{k}}$ pair, $m_i$ is the mass of the $i$-th quark,
and $\veck = \vecp_j+\vecp_{\overline{k}}$.
The matrix elements of $(\lambda_i\cdot \lambda_{j\overline{k}})$ and
$\sigma_{j\overline{k}}$ are given in appendix A.

The single baryon wave function, $\phi_B$, which is antisymmetrized in terms 
of quark exchanges is given by a product of orbital, 
color and spin-flavor parts:
\[
\phi_B(\xbld{\xi}_B)=\varphi_B(\xbld{\xi}_B)C([111]_c)SF([3]_{\sigma f}).
\]
The meson wave function, $\phi_M$, is also given by a product of orbital, 
color (singlet), spin, and flavor parts:
\[
\phi_M(\xbld{\xi}_M)=\varphi_M(\xbld{\xi}_M)C({\rm singlet}) S F.
\]
We employ the gaussian form for these single baryon and meson wave functions 
$\varphi (\xbld{\xi})$. 
\begin{eqnarray*}
\varphi_B(\xbld{\xi}_B)&=&g(\xbld{\xi}_1,\sqrt{2}b)g(\xbld{\xi}_2,
\sqrt{\frac{3}{2}}b) \\
\varphi_M(\xbld{\xi}_M)&=&g(\xbld{\xi}_M,\sqrt{2}b)
,
\end{eqnarray*}
where
$\xbld{\xi}_B$ and $\xbld{\xi}_M$ are internal coordinates of baryon $B$ and meson $M$. 
The $\xbld{\xi}_B$ and $\xbld{\xi}_M$ are given by the i-th quark 
coordinates $\xbld{r}_i$ as follows:
\begin{eqnarray*}
\xbld{\xi}_B &=& (\xbld{\xi}_{1},\xbld{\xi}_{2})=(\xbld{r}_1-\xbld{r}_2,
\frac{\xbld{r}_1+\xbld{r}_2}{2}-\xbld{r}_3),\\
\xbld{\xi}_M &=& (\xbld{r}_4-\xbld{r}_5).
\end{eqnarray*}
The $g(\xbld{r},b)$ is a normalized gaussian wave function with a size parameter $b$  
given by
\[
g(\xbld{r},b)=(\sqrt{\pi} b)^{-3/2} \exp(-\frac{r^2}{2b^2}).
\]

The parameter used in this work are shown in Table I.
The calculated baryon and meson masses employing the above hamiltonian and wave functions
 are given in Table II. 
We assume the ideal mixing for $\eta$ in the coupled channel QCM, because
the channel $\Lambda\eta$ is not open in the concerning energy region, and the 
u and d components must be important. 
The masses of $\eta_{ud}$ = $({\rm u}\overline{\rm u}+{\rm d}\overline{\rm d})/\sqrt{2}$
and $\eta_{s}$ = ${\rm s}\overline{\rm s}$
are also be shown in Table II.
 
%
%
\begin{table*}[tb]
\caption{Model parameters.}
\begin{center}
\label{tbl:QMparam}
\def\SPA{\phantom{0}}
\def\SPB{\phantom{$-$0}}
\def\SPC{\phantom{.0}}
\tabcolsep=0.5mm
\begin{tabular}{ccccccccccccccc}
\hline 
$m_{\rm u}$ &$m_{\rm s}$ 
& $\aconf$ & $\alpha_s$ &$\aiiit$ & $v_0$ & $V_0$ & $b$  \\
(MeV)&(MeV)&(MeV/fm)& &(MeV fm$^3$)  & (MeV fm$^3$)& (MeV) & (fm)  \\
313 & 507 & 163.33 & 0.5040 & $-$94.10  & 9.7 & $-$468.49 & 0.49 \\ \hline
%
%
\end{tabular}
\end{center}
\end{table*}
%
%
\begin{table}[btp]
\caption{Masses of baryons and mesons given by the present model.
All entries are given in MeV.}
\label{tbl:mass}
\begin{center}
\def\SPC{\phantom{.0}}
\begin{tabular}{l@{~~}ccccccc@{~~}ccccc}
\toprule
& N\SPC & $\Sigma$\SPC & $\Xi$\SPC & $\Lambda$\SPC & $\pi$\SPC & K\SPC & $\eta$\SPC  & $\eta'$\SPC& $\eta_{ud}$\SPC & $\eta_s$\SPC\\
%
Model& 939\SPC & 1188\SPC & 1308\SPC & 1113\SPC & 139\SPC & 487\SPC &\SPC&\SPC& 552 & 931 \\ 
Exp.\cite{pdg}\ 
& 938.9 & 1190.5 & 1318.1 & 1115.7 & 138.0 & 495.0 & 547.8 & 957.8 & - & -
\\ \hline
\end{tabular}
\end{center}
\end{table}

\section{Quark cluster model}
Here, we briefly explain the quark cluster model.  
A baryon consists of three quarks and a meson consists of quark and an antiquark,
whose effective mass is about 
300 MeV. The baryon-meson system, which consists of 4 quarks and 1 antiquark, is written 
as a product of these single baryon and meson wave functions 
$\phi_B(\xbld{\xi}_B)$ and $\phi_M(\xbld{\xi}_M)$  and the 
relative wave function
$\chi(\xbld{R}_{BM})$ between these two clusters:
\begin{equation}
\label{relwav}
\Psi(\xbld{\xi}_B ,\xbld{\xi}_M, { \xbld{R}}_{BM})={\cal A}[\phi_B(\xbld{\xi}_B)
\phi_M(\xbld{\xi}_M)\chi({\xbld{R}}_{BM})].
\end{equation}
The ${\cal A}$ is an antisymmetrization operator among 4 quarks,
\[
{\cal A}=1-3P_{34}.
\]
Here we assume that the baryon wave function is antisymmetrized.
 
Employing the known baryon and meson wave functions, $\phi_B(\xbld{\xi}_B)$ and $\phi_M(\xbld{\xi}_M)$, 
we obtain the following equation (RGM equation) to determine 
the relative wave function
$\chi (\xbld{R}_{BM})$:
\begin{equation}
\int [H_{RGM}(\xbld{R},\xbld{R}')-
EN_{RGM}(\xbld{R},\xbld{R}')]\chi(\xbld{R}')d\xbld{R}'=0,\label{rgm}
\end{equation}
where the hamiltonian $H_{RGM}$ and normalization $N_{RGM}$ kernels 
are given by
\begin{eqnarray*}
&{}& \left \{ \begin{array}{c}
H_{RGM}(\xbld{R},\xbld{R}')\\
N_{RGM}(\xbld{R},\xbld{R}')
\end{array} \right \}
=\int \phi^\dagger_B(\xbld{\xi}_B)
\phi^\dagger_M(\xbld{\xi}_M)\delta(\xbld{R}-\xbld{R}_{BM})
\left\{ \begin{array}{c}
H\\ 1
\end{array} \right\} \\
&{}& \times
{\cal A}[\phi_B(\xbld{\xi}_B)\phi_M(\xbld{\xi}_M)\delta(\xbld{R}'-\xbld{R}_{BM})]d\xbld{\xi}_Bd\xbld{\xi}_Md\xbld{R}_{BM}.
\end{eqnarray*}
 Here, $H$ is the total hamiltonian of the system where the center
of mass kinetic energy is subtracted.

In the following, we explain how to modify the relative kinetic energy for the baryon-meson 
scattering problem. The RGM kernel $H$ and $N$ in eq.\ (\ref{rgm}) consist of direct 
and exchange terms.
We take out the direct term of the kinetic energy operator $K_{D}$ in the RGM kernel $H$.
By subtracting the internal kinetic energy with the direct norm kernel $N_{D}$, 
the RGM kernel of the relative kinetic energy $K_{R}(\xbld{R},\xbld{R})$ is given by 
\[
K_{R}(\xbld{R},\xbld{R})=K_{D}(\xbld{R},\xbld{R})
 - 3K_0 \times N_{D}(\xbld{R},\xbld{R}), \hspace{10pt}\mbox{where} 
\hspace{10pt}K_0=\frac{3}{4m_q b^2} .
\]
Then the relative kinetic energy is modified 
in the following way.
\[
K_{R}(\xbld{R},\xbld{R}) \rightarrow K_{R}(\xbld{R},\xbld{R})
 \times \frac{6 m_q}{5} \frac{1}{\mu},
\]
where $\mu$ is the reduced mass for the baryon-meson system calculated by using the masses 
given in Table II.

The extension to the coupled-channel calculation is straightforward. 
The inclusion of the $\Lambda^1$ pole as a bound state embedded in the continuum  
is also straightforward.
 We add  the pole  at a certain energy 
to the baryon-meson wave function.
\[ 
\Psi=\Psi(\xbld{\xi}_B ,\xbld{\xi}_M, { \xbld{R}}_{BM}) + \Psi(q^3).
\]
The coupling of the baryon-meson state with the $\Lambda^1$ pole is 
treated in the following way.
First we calculate $\bra 0s^20p |V_{3q-5q} | 0s^5\ket $ 
with $V_{3q-5q}$ given in eq.(\ref{eq:V3q5q}).
This calculation can be done straightforwardly (see appendix A).
Their values calculated by the present parameter set are listed in Table \ref{tbl:VtrValue}. 
Note that the $|0s^5\ket $ corresponds to a state where the relative wave function 
$\chi (\xbld{R}_{BM})$ is a gaussian
with the size parameter $B=\sqrt{5/6}b$ as follows: 
\[
\chi(\xbld{R}_{BM})=g(\xbld{R}_{BM},B=\sqrt{\frac{5}{6}}b).
\]
The dependence on the relative distance $\xbld{R}_{BM}$ of the baryon-meson coupling is 
taken into account by multiplying an overlap with the above $|0s^5\ket $ state. This 
overlap is given by the norm kernel of the generator coordinate method (GCM), 
whose details are given in appendix B.       

\begin{table}[btp]
\caption{Values of the transfer matrix elements
between the q$^3$ flavor-singlet spin-${1\over 2}$ $(0s)^20p$ state ($\Lambda^1$)
and the q$^4\qbar(0s)^5$ baryon-meson states, as well as those between 
the q$^3$ isospin-1 spin-${1\over 2}$ $(0s)^20p$ state ($\Sigma^*$) and the baryon meson states.
All entries are given in MeV.}
\label{tbl:VtrValue}
\begin{center}
\def\SPC{\phantom{.0}}
\begin{tabular}{ccc@{~~}cccc@{~~}ccccc}
\toprule
 $\bra \Lambda^1| V^{tr}| \Sigma\pi \ket$ 
& $\bra \Lambda^1| V^{tr}| {\rm N}\overline{\rm K} \ket$ 
& $\bra \Lambda^1| V^{tr}| \Lambda\eta_{ud} \ket$ 
& $\bra \Sigma^*| V^{tr}| \Lambda\pi \ket$ 
& $\bra \Sigma^*| V^{tr}| \Sigma\pi \ket$ 
& $\bra \Sigma^*| V^{tr}| {\rm N}\overline{\rm K} \ket$ 
\\
%
140& $-$85 & 53 & $-$32 & $-$51 &60
\\ \hline
\end{tabular}
\end{center}
\end{table}

\section{Results and Discussion}
In describing the $\Lambda$(1405), we first consider the following  SU(3) octet baryon and meson 
systems.
\[
\xbld{8}_B \times \xbld{8}_M = \xbld{1}_{BM}+\xbld{8}_{BM}+\xbld{8}_{BM}+\xbld{10}_{BM}+
\overline{\xbld{10}}_{BM}+\xbld{27}_{BM}.
\]
The strangeness=$-$1 and isospin $T=0$ state appears in the 
$\xbld{1}_{BM},\xbld{8}_{BM}$ and $\xbld{27}_{BM}$ states. 
These four states are given by a linear 
combination of the following four baryon-meson systems.
\[
\Sigma \pi,\hspace{5pt}\rmN \Kbar,\hspace{5pt} \Lambda \eta,\hspace{5pt} \Xi K.
\]
For example, the flavor singlet state, $|\xbld{1}_{BM}\ket$, is given by 
\[
|\xbld{1}_{BM}\ket = \sqrt{\frac{3}{8}}|\Sigma \pi\ket -\frac{1}{2}|\rmN \Kbar\ket +
\sqrt{\frac{1}{8}}|\Lambda \eta\ket +
\frac{1}{2}|\Xi K\ket .
\]

First we estimate the contributions from the color-spin part of the OGE, 
usually called as the color magnetic interaction (CMI), 
which plays the most important role in the 
constituent quark model.
The matrix elements for the SU(3) octet baryon is 
\[
\bra -\sum_{i<j}^3 \xbld{\sigma}_i \cdot \xbld{\sigma}_j \lambda_i \cdot \lambda_j\ket =-8
\hspace{10pt} \mbox{for the octet baryons}.
\]
Because the matrix element for the SU(3) decuplet baryon is 8, this color-spin interaction  
 gives the main contribution to the mass difference between the octet and decuplet baryons
 in the OGE model.  
For the SU(3) octet pseudoscalar mesons, the matrix element is given by
\[
\bra \xbld{\sigma}_4 \cdot \xbld{\sigma}_5 \lambda_4 \cdot \lambda_5^{\ast}\ket =-16
\hspace{10pt} \mbox{for the octet pseudoscalar mesons}.
\]
The matrix element for the SU(3) octet vector meson is 16/3. Therefore the CMI 
again succeeds to give the larger contribution to the mass difference between 
pseudoscalar  and vector mesons than that of the octet and decuplet baryons.

Now let us discuss the interaction due to the color magnetic term of OGE between 
the baryon 
and meson system given as
\[
-\sum_{i<j}^4 \xbld{\sigma}_i \cdot \xbld{\sigma}_j \lambda_i \cdot \lambda_j
+\sum_{i=1}^4 \xbld{\sigma}_i \cdot \xbld{\sigma}_5 \lambda_i \cdot \lambda_5^{\ast}
-(-18-16).
\]
 It is obvious that the direct term has no contribution to the baryon meson 
interaction because each of the baryon and meson is color-singlet. 
There is, however, a sizable contribution 
when we include the exchange term due to the antisymmetrization over the quarks. 
The results for the isospin $T=0$ baryon-meson system are shown in Table \ref{tbl:llss}.

\begin{table*}[tb]
\caption{Matrix elements of the color magnetic operator for $T=0$.}
\renewcommand{\arraystretch}{2.2}
\begin{tabular}{|c|c|c|c|c|} \hline
 & $\Sigma \pi$ & N$\Kbar$ & $\Lambda \eta$ & $\Xi K$ \\ \hline
 $\Sigma \pi$ & $\displaystyle -\frac{16}{3}$ & $\displaystyle \frac{116\sqrt{7}}{21}$ &
 $\displaystyle -\frac{16\sqrt{105}}{105}$ & 0 \\ \hline
 N$\Kbar$ & & 0 & $\displaystyle \frac{28\sqrt{15}}{15}$ & 0   \\ \hline
 $\Lambda \eta$ & & & $\displaystyle \frac{112}{15}$ & $\displaystyle -\frac{40\sqrt{70}}{21}$ \\ \hline
$\Xi K$  & & & & $\displaystyle -\frac{160}{21}$ \\ \hline
\end{tabular}
\label{tbl:llss}
\end{table*}

In order to compare the features of this OGE model to those of the baryon meson picture,
 we also show  the matrix elements which appear 
in the Weinberg-Tomozawa term in the chiral unitary model
 in Table \ref{tbl:ff}; they are proportional to the inner product of 
 the flavor-SU(3) generator $F$, ($\sum_{i<j}F_i \cdot F_j$).
As seen in these tables, the $\Sigma \pi$ channel has a strong attraction in both 
of the interactions. 
The N$\Kbar$ channel, however, has a strong attraction only 
in the above \FF-type interaction;
the contribution from the CMI to the diagonal N$\Kbar$ interaction vanishes.
Therefore, as long as we restrict ourselves to the OGE model with the q$^4\qbar$\ system, 
it is rather difficult to explain 
the $\Lambda$(1405) peak as a N$\Kbar$ bound state 
appearing in the $\Sigma\pi$ scattering.

\begin{table*}[tb]
\caption{Matrix elements of \FF\ operator for $T=0$.}
\begin{tabular}{|c|c|c|c|c|} \hline
 & $\Sigma \pi$ & N$\Kbar$ & $\Lambda \eta$ & $\Xi K$ \\ \hline
 $\Sigma \pi$ & $-$8 & $\sqrt{6}$ &
 0 & $-\sqrt{6}$ \\ \hline
 N$\Kbar$ & & $-$6 & $3\sqrt{2}$ & 0   \\ \hline
 $\Lambda \eta$ & & & 0 & $-3\sqrt{2}$ \\ \hline
$\Xi K$  & & & & $-$6 \\ \hline
\end{tabular}
\label{tbl:ff}
\end{table*}
\bigskip

Now let us show the results of the dynamical calculation.
Here we consider 
$\Sigma\pi$, N$\Kbar$ and $\Lambda\eta_{ud}$ channels for isospin $T=0$ state. 
The effect of each channel couplings will also be discussed.   

In the following figures,  we show the phase shift, $\delta$, 
of the  $\Sigma\pi$ 
channel for the angular momentum $L=0$. 
We also show the $\Sigma\pi$ mass spectrum given by
\[
|1-\eta {\rm e}^{2{\rm i}\delta}|^2/k,
\]
where scattering matrix $S$ is denoted as $ S=\eta {\rm e}^{2{\rm i}\delta}$ and $k$ is the 
relative wave number for the $\Sigma\pi$ scattering.

\begin{figure}[tbp]
\caption{Phase shift and the mass spectrum
of the  $\Sigma\pi$ 
channel for the angular momentum $L=0$ given  by  the QCM calculation 
without the $\Lambda^1$ pole 
with the constituent-quark reduced mass, $6 m_q/5$. }
\begin{center}
\includegraphics[height=3.5in]{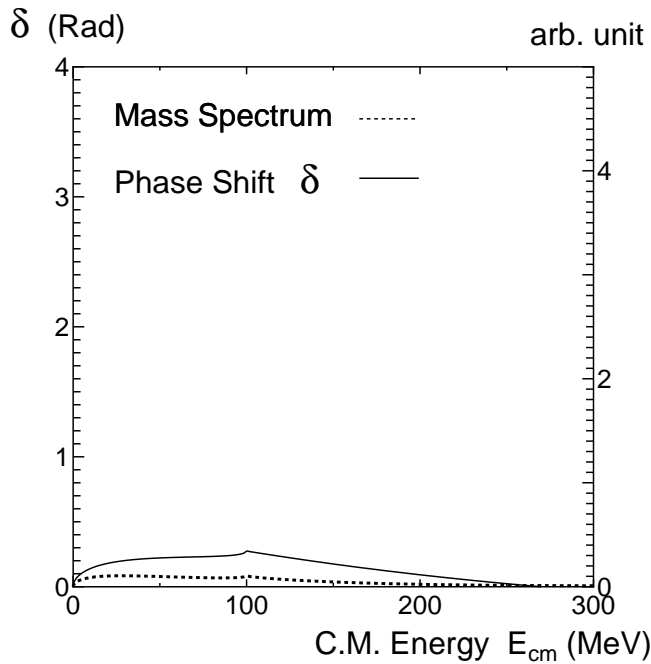}
\end{center}
\label{fig:massQMC}
\end{figure}

First we show the results of the pure QCM calculation without the $\Lambda^1$ pole 
in Fig.\ \ref{fig:massQMC},
where the reduced mass for the relative 
kinetic energy is  $6 m_q/5$. 
As seen in the figure, there is no structure in the mass spectrum 
when the 
the coupling to the $\Lambda^1$ pole is not included. 
   
\begin{figure}[tbp]
\caption{Phase shift and the mass spectrum
of the  $\Sigma\pi$ 
channel for the angular momentum $L=0$ given  by  the QCM calculation 
with the $\Lambda^1$ pole at $\E0pole$ = 160 MeV and
 the constituent-quark reduced mass, $6 m_q/5$. }
\begin{center}
\includegraphics[height=3.5in]{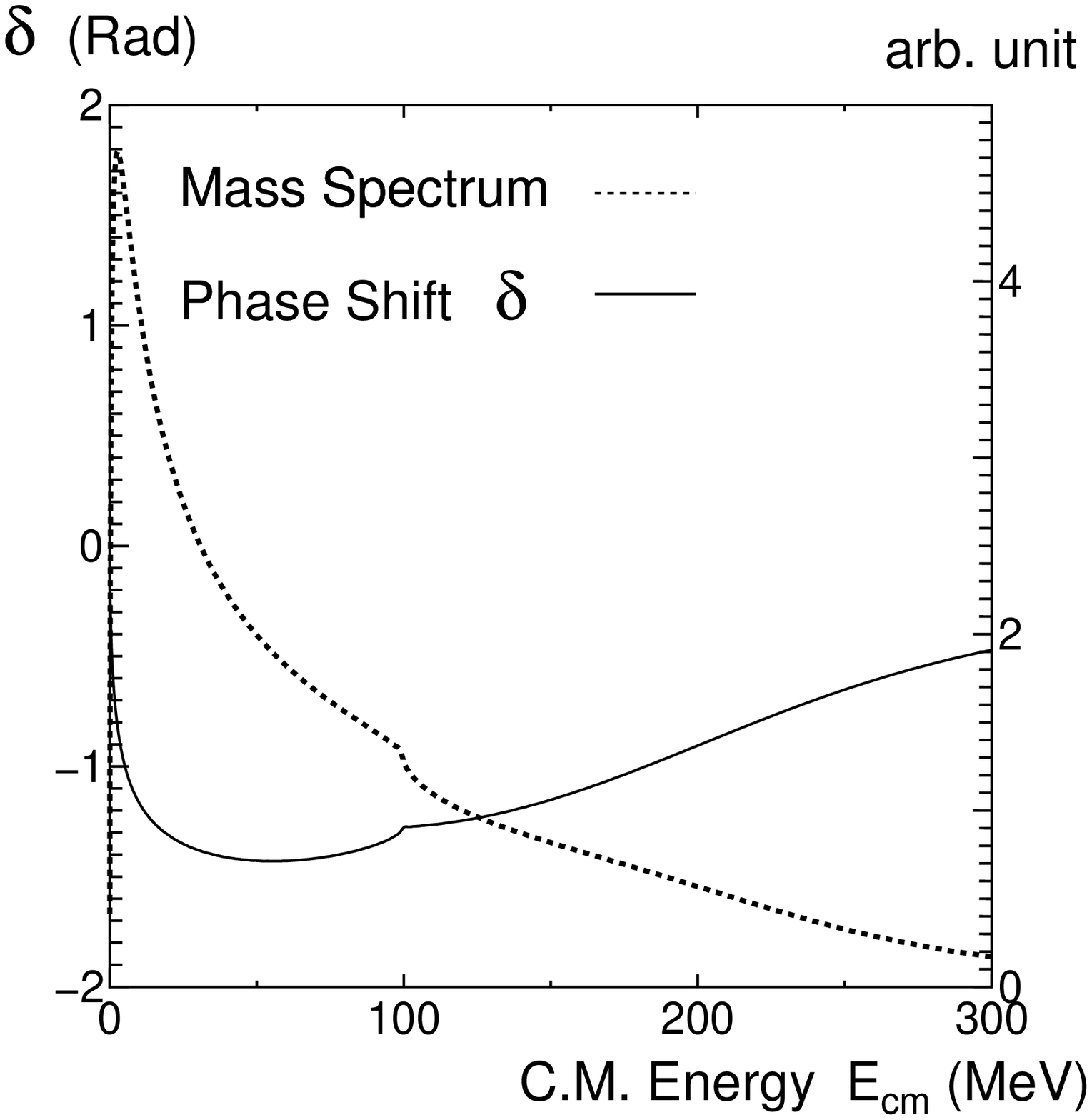}
\end{center}
\label{fig:massQMCQ}
\end{figure}

Next we show the result of the calculation including a coupling with the $\Lambda^1$ pole 
in the baryon-meson 
scattering in Fig.\ \ref{fig:massQMCQ}. 
Before the coupling to the baryon-meson is switched on,
the energy of the pole, $\E0pole$, 
is assumed to be 160 MeV above 
the $\Sigma\pi$ threshold.
As seen in the figure, there appears a bound state with the binding energy 1.7 MeV.
The probability of the $\Lambda^1$ pole in this bound state is 0.20, 
while those of $\Sigma\pi$, N$\Kbar$ and $\Lambda\eta_{ud}$ are
0.77, 0.04, and 0.004, respectively.
This state is essentially a $\Sigma\pi$ state bound by the strongly attractive 
CMI with the help of the $\Lambda^1$ pole coupling. 

The phase shift at the low energy 
region  for the $\Sigma\pi$ channel becomes negative, because there is
a bound state.
The phase shift, however, goes eventually  to zero, rather than $-\pi$, because we
put an extra closed state, the $\Lambda^1$ pole, in the system.
When we move the original energy of the pole higher, {\it e.g.}\ 200 MeV above the
$\Sigma\pi$ threshold, the bound state disappears and becomes a resonance just above the 
$\Sigma\pi$ threshold. 
Model ambiguity allows us to change $\E0pole$ higher,
or to make the coupling to the pole smaller. Then, however, the resonance
just vanishes without moving to 
the higher energy: namely, no resonance appears around $E$ = 1405 MeV by this model.
It is because the  attraction in the $\Sigma\pi$ channel is much stronger than
that of the N$\Kbar$ channel.

 We must note, however, no peak appears around the observed 
energy region either even if we employ the \FF-type interaction, so long as the 
the rather heavy reduced mass, $6 m_q/5=376$ MeV, is used for the $\Sigma\pi$ channel.
There appears a $\Sigma\pi$ bound state due to the heavy reduced mass 
when the attraction for $\Sigma\pi$ and N$\Kbar$ channels is strong enough. 
It is, therefore, 
important to use the observed value, 124 MeV for $\Sigma\pi$ and 324 MeV for N$\Kbar$,
which add a repulsive effect especially to the
$\Sigma\pi$ channel.
Therefore, in the following calculations, we employ the realistic reduced 
masses, calculated from the masses given by the present quark model (Table \ref{tbl:mass}). 

\begin{figure}[tbp]
\caption{Phase shift and the mass spectrum
of the  $\Sigma\pi$ 
channel for the angular momentum $L=0$ and the wave function at the resonance.
The QCM calculation is performed 
with the $\Lambda^1$ pole at $\E0pole$ = 160 MeV and
 the realistic reduced mass.}
\begin{center}
\includegraphics[height=3.5in]{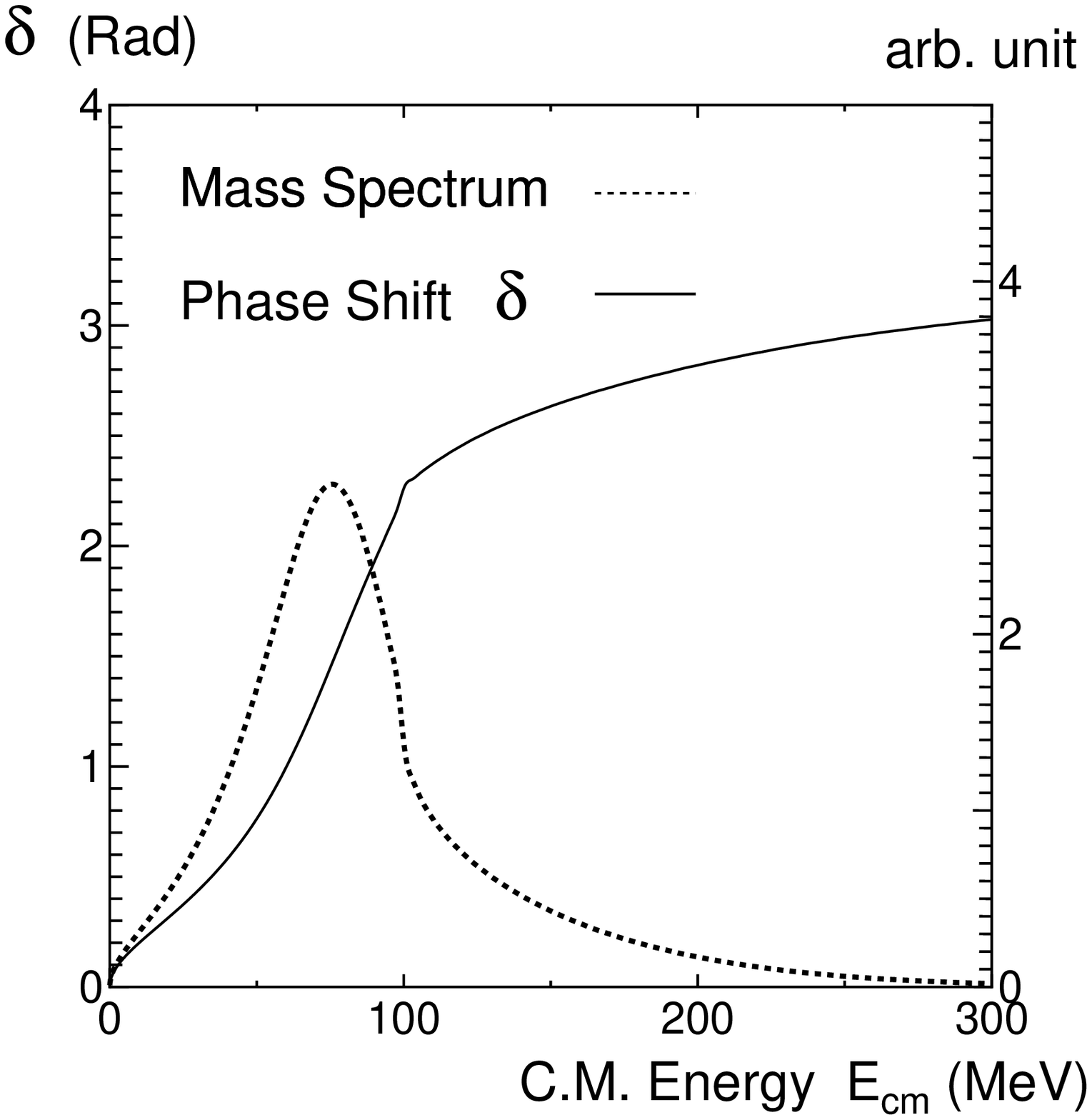}%
\includegraphics[height=3.2in]{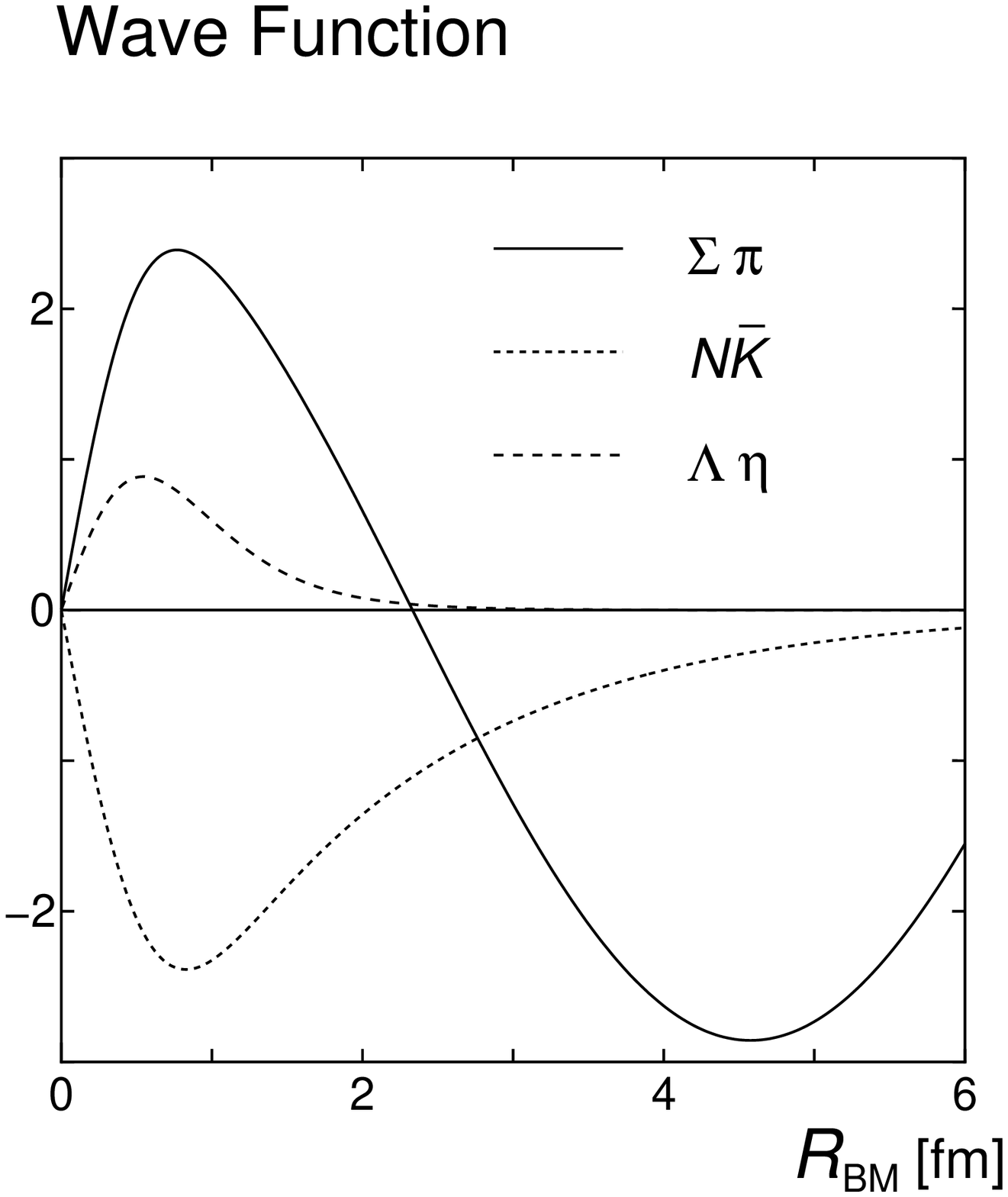}
\end{center}
\label{fig:mass3CQ}
\end{figure}

Now we show the results of the coupled channels calculation employing 
the realistic reduced  masses
including the $\Lambda^1$ pole (Fig.\ \ref{fig:mass3CQ}). Here it is also assumed 
that $\E0pole$ is 160 MeV 
above the $\Sigma\pi$ threshold. 
As seen in the figure, the bound state found in the previous calculation vanishes.
Instead, an resonance appears around 75 MeV above the $\Sigma\pi$ threshold,
which corresponds to 1404 MeV.
The width of this peak is found to be 55 MeV.
Both of the results agree with the observed mass and width of $\Lambda$(1405):
1406$\pm$4 MeV and 50$\pm$2 MeV, respectively \cite{pdg}.

The wave functions at the resonance ($k=0.70{\rm fm}^{-1}$) are also shown in 
Fig.\ \ref{fig:mass3CQ}.
The ratio of the mixing probabilities of the $\Lambda^1$ pole and N$\Kbar$ state is roughly 2.8,
which indicates that the resonance is essentially the $\Lambda^1$ pole.
The coupling to the baryon-meson channels, however, 
is very important to reproduce the resonance at such a low energy: it causes the energy 
shift of 85 MeV, which lowers the pole position from 160 MeV down to 75 MeV. 

\begin{figure}[tbp]
\caption{The diagonal parts of the S-matrix for the coupled channel baryon-meson scattering.}
\begin{center}
\includegraphics[height=3.5in]{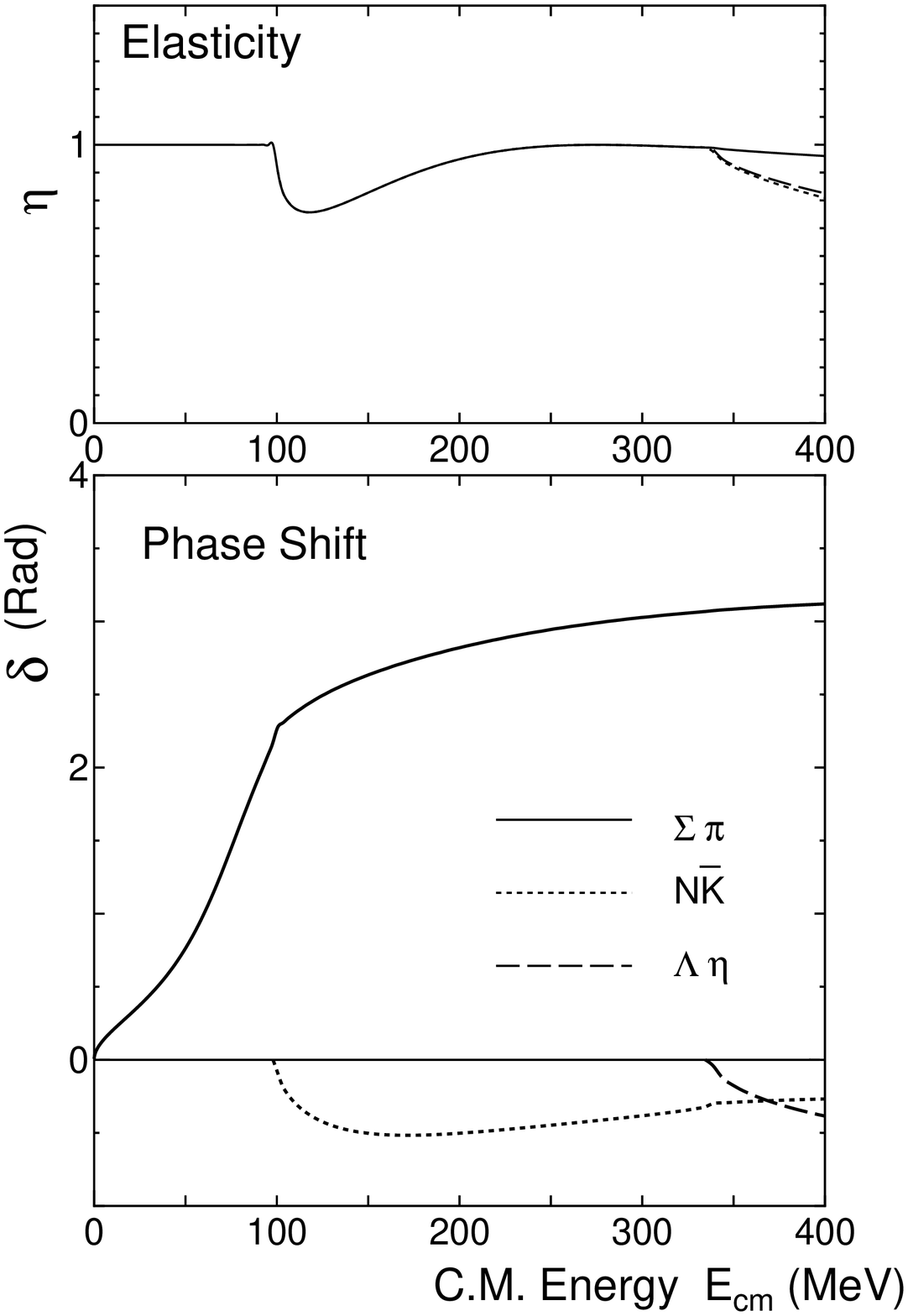}
\end{center}
\label{fig:Smat3CQ}
\end{figure}

The diagonal parts of the S-matrix for the coupled channel baryon-meson scattering are 
shown in  Fig.\ \ref{fig:Smat3CQ}. 
Above the N$\Kbar$ threshold, the elasticity decreases sharply.
The N$\Kbar$ phase shift goes to negative: the scattering length is $-0.75+i0.38$ fm, 
which, as a simple model,  agrees 
with the experimental value, $(-1.70\pm 0.07) +i(0.68\pm 0.04)$ fm 
\cite{Martin:1980qe}, reasonably well.

\begin{figure}[tbp]
\caption{The effect of the channel couplings on the peak.}
\begin{center}
\includegraphics[height=3.5in]{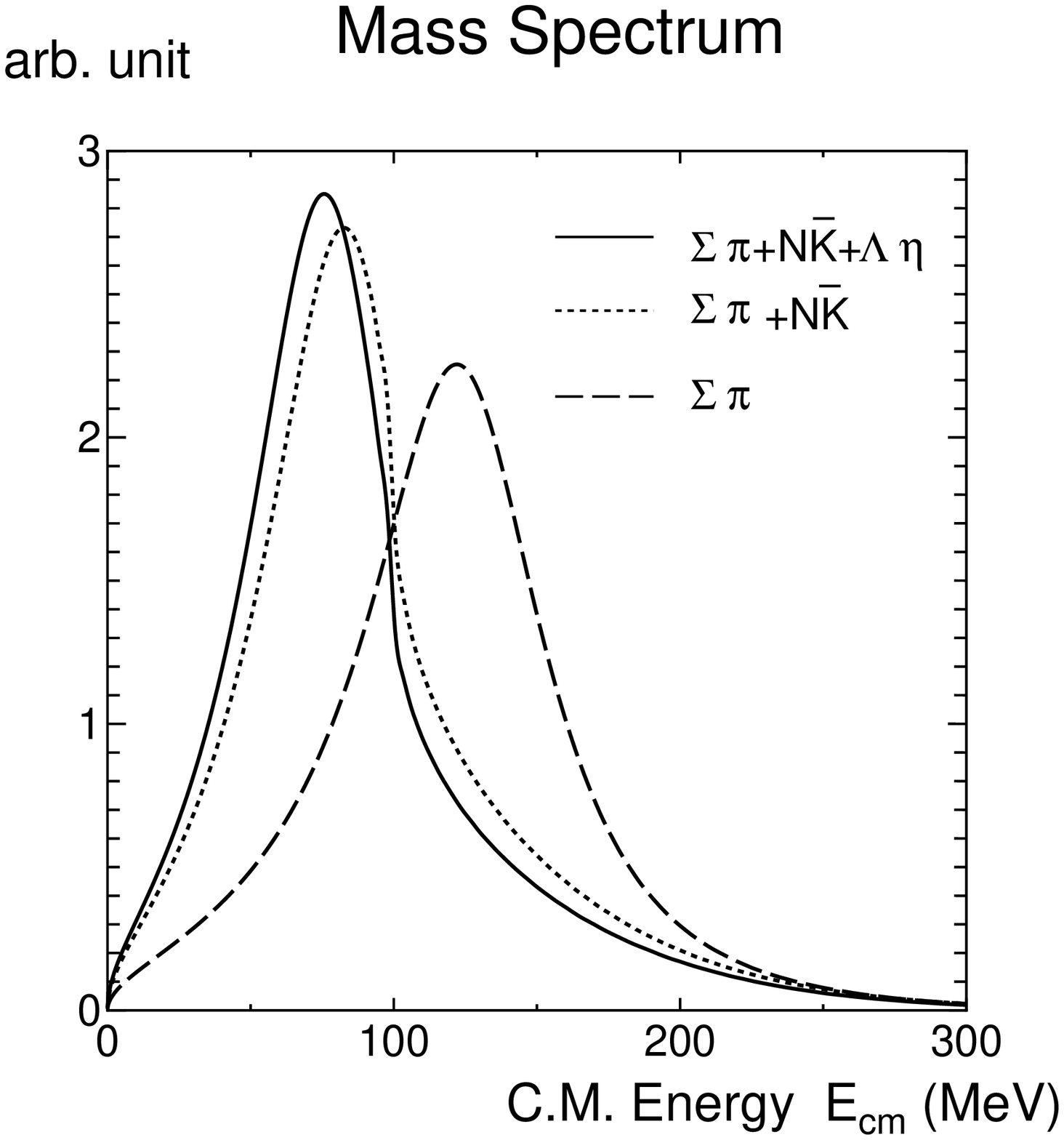}
\end{center}
\label{fig:massCQ321}
\end{figure}

Next we discuss the effect of the channel couplings. 
The mass spectra given by the single channel $\Sigma\pi$ calculation, the coupled channel 
calculations with two channels 
$\Sigma\pi$+N$\Kbar$ and with three channels $\Sigma\pi$+N$\Kbar$+$\Lambda\eta_{ud}$ 
are shown in Fig.\ref{fig:massCQ321}. As seen in the figure, in the single baryon-meson channel
calculation, there appears a resonance around 
122 MeV above the $\Sigma\pi$ threshold, 
which means that the resonance is shifted by 38 MeV by the coupling.
Then the peak is shifted lower down to 84 MeV  above the $\Sigma\pi$ threshold by adding the 
N$\Kbar$ channel. The inclusion of the  $\Lambda\eta_{ud}$ channel shifts the resonance 
a little lower
by 9 MeV from the one given by the two-channel calculation.
\bigskip

The calculation based on the q$^4\overline{\rm q}(0s)^5$ configuration
shows that some of the flavor-octet negative-parity baryons, {\it e.g.} $\Sigma^* ({1\over 2}^-)$, 
also  gains strong attraction \cite{Hogaasen:1978jw}.
In the present dynamical model, however, no bound state nor resonance is found
without the pole in 
the $\Lambda \pi$-$\Sigma \pi$-N$\overline{\rm K}$ ($TJ$)=(1$\half$) scattering; 
the situation is similar to the 
the above ($TJ$)=(0$\half$) case.
The q$^3$ $(0s)^20p$ state also exists in this channel, but has much heavier mass: 
128 MeV above the $\Lambda^1$ mass if the mass difference is calculated by the conventional quark model. 
Also, the matrix elements between the baryon-meson states and
the $\Sigma^*({1\over 2}^-)$ $(0s)^20p$ pole are much smaller (Table \ref{tbl:VtrValue}).
We perform the coupled-channel QCM calculation also for $T=1$ channel with this pole at 360 MeV above the 
$\Lambda\pi$ threshold, which corresponds to 125 MeV above the $\Lambda^1$ pole used for the T=0 channel.
A resonance appears in the N${\overline{\rm K}}$ channel when the coupling to this pole
is switched on.  The energy of the resonance, however, is shifted only by 12 MeV:
the peak appears at about the same energy as that of the pole.
Namely, in the present model scheme, no resonance is found in this channel at an energy as low as $\Lambda(1405)$.

As we mentioned before, 
we assume that $\E0pole$
is 160 MeV above the $\Sigma\pi$ threshold,
which corresponds to 1489 MeV.
We choose this value so that the system produces the resonance around 1405 MeV
in the present model scheme.
The value, 1489 MeV, is not unreasonable because the predicted value in ref.\ \cite{Isgur:1978xj} is 1490 MeV.
In this scheme, however, it is somewhat lower than the energy where the pole is considered to exist.
For example, the mass of the flavor-singlet q$^3$ $(0s)^20p$ state calculated with the present parameter set is 
1551 MeV. 
Moreover, $\Lambda$(1520) should be closer to the mass of the $\Lambda^1$ pole because
it couples to the scattering states of the ground state baryons and mesons
only by the non-central part of the interaction.
Thus, the pole probably exist at around 1520-1550 MeV in this scheme.
Nevertheless, we consider it very interesting 
that the present simple model with the assumption that OGE induces
the coupling only between the q$^4\overline{\rm q}(0s)^5$ and q$^3(0s)^20p$ state
can give such a peak.
We argue that the mechanism of the $\Lambda$(1405) resonance can be the baryon-meson scattering
coupled to the three-quark state.

\section{Summary}

We have investigated the negative-parity $\Lambda$(1405) state 
in terms of the baryon-meson $S$-wave scattering by employing a quark cluster model.
The model hamiltonian
has qq and q$\qbar$ hyperfine interactions coming from  the one-gluon-exchange (OGE)
as well as the instanton-induced interaction ($\Viii$). 
The parameters are taken so that all the masses of the 
octet baryons and mesons are reproduced quite well. 

We perform the $\Sigma\pi$-N$\Kbar$-$\Lambda\eta_{ud}$ coupled channel QCM calculation
with and without the 
coupling to the $\Lambda^1$ pole by OGE.
The results show that 
(1) there is a strong attraction in the $\Sigma\pi$ channel but not in the N$\Kbar$ channel,
(2) no peak is found in the $\Sigma\pi$-N$\Kbar$-$\Lambda\eta_{ud}$ coupled channel QCM calculation
if we employs the realistic reduced mass for the kinetic energy,
and (3) a reasonable peak appears if the $\Lambda^1$ pole is included above the N$\Kbar$ threshold.

In the baryon-meson picture where the flavor-flavor type interaction, \FF,  is employed, 
the peak appears because of the 
 attraction in the N$\Kbar$ channel\cite{Oset:1997it}. 
In the present scheme of the color-spin interaction, there is no such an attraction in the N$\Kbar$ channel,
which requires the introducing the $\Lambda^1$ pole to reproduce the resonance.
The spin-independent \FF-type interaction appears when we introduce the vector meson exchange potential 
between quarks in addition to the Goldstone boson exchange or to the gluonic interactions. 
To include such an interaction may be interesting but beyond the scope of the present work.
Here we would like to emphasize that 
a quark model with the one-gluon exchange and the instanton-induced interaction
can reproduce the bulk feature of the $\Lambda(1405)$ with the help of the $\Lambda^1$ pole.

\begin{acknowledgements}
This work is supported in part by a Grant-in-Aid for Scientific Research
from JSPS (Nos.\ 
15540289, 
17540264, 
and 18042007
).
\end{acknowledgements}

\def\ubari{\overline{u}}
\def\sbari{\overline{s}}
\def\rmu{{\rm u}}
\def\rms{{\rm s}}
\def\lamjkbar{{\lambda_{\overline{k}j}}}
\def\sigjkbar{{\bm{\sigma}_{\overline{k}j}}}
\appendix

\section{Coupling between the $\Lambda^1$-pole and the Baryon-meson state}
We assume that OGE induces the coupling between the 
spin-1/2 $(0s)^2(0p)$
state, and q$^4\qbar(0s)^5$ state.

First let us calculate the matrix element between u and us$\sbar$ states (See Fig.\ \ref{fig:diag1}).
Each of these states can be written as
\begin{eqnarray}
\bra \rmu(\xi'_i)| &=& \bra 0| a_u(\xi'_i)
\\
|\rmu(\xi_i)\rms(\xi_j)\sbar(\xi_k)\ket &=& a_u^\dag(\xi_i)a_s^\dag(\xi_j)b_{\sbari}^\dag(\xi_k)|0\ket~.
\end{eqnarray}
Then the matrix element is given by
\begin{eqnarray}
M = \alpha_s {1\over \sqrt{2\epsilon'\!{}_i\;2\epsilon_i\;2\epsilon_j\;2\epsilon_k}}
(\ubari_{\xi'_i}{\cal O}^a_\mu u_{\xi_i}) \;{\cal D}_{\mu\nu}\;
(\sbari_{-\xi_k}{\cal O}^a_\nu s_{\xi_j}) \;e^{i(\bm{p}'\!\!{}_i-\bm{p}_i-\bm{p}_j-\bm{p}_k)\bm{x}},
\end{eqnarray}
where $u_\xi$ [$s_\xi$] is a spinor of the u-quark [s-quark]
with the quantum number $\xi$, and $\sbar_{-\xi}$ is a spinor of the s-antiquark with 
quantum number $\xi$.
Since the vertex and the lowest-order term of the propagator of the gluon are given by
\begin{eqnarray}
{\cal O}^a_\mu &=& g{\lambda^a\gamma_\mu\over 2}\\
{\cal D}_{\mu\nu} &=& {\pi \over m^2}g_{\mu\nu}~,
\end{eqnarray}
the matrix element becomes
\begin{eqnarray}
M = \alpha_s {\pi \over m_s^2} {1\over \sqrt{2\epsilon'\!{}_i\;2\epsilon_i\;2\epsilon_j\;2\epsilon_k}}
e^{i(\bm{p}'\!\!{}_i-\bm{p}_i-\bm{p}_j-\bm{p}_k)\bm{x}}
 \Big\{
(\ubari_{\xi'\!{}_i}{\lambda^a{\gamma^0} \over 2}u_{\xi_i})(\sbari_{-\xi_k}{\lambda^a{\gamma_0} \over 2} s_{\xi_j})
-(\ubari_{\xi'\!{}_i}{\lambda^a\bm{\gamma}\over 2}\, u_{\xi_i})(\sbari_{-\xi_k}{\lambda^a\bm{\gamma}\over 2}\, s_{\xi_j})
\Big\}.
\end{eqnarray}

After the nonrelativistic reduction, the lowest-order term of the $(p/m)$ expansion becomes
\begin{eqnarray}
M&=& (\lambda_i\cdot\lamjkbar) {\alpha_s\over 4} {\pi \over m_s^2} 
e^{i(\bm{p}'\!\!{}_i-\bm{p}_i-\bm{p}_j-\bm{p}_k)\bm{x}}
 \Big\{
\big(w_{u\,\xi'\!{}_i}^*~w_{u\,\xi_i}\big)
\big(w_{s\,-\xi_k}^*{1\over 2 m_s}\bm{\sigma}\cdot(\vecp_j\!+\vecp_k)~w_{s\,\xi_j}\big)
\nonumber\\
&&-\big(w_{u\,\xi'\!{}_i}^*(\vecp'_i\!+\vecp_i+i\bm{\sigma}\!\times\!(\vecp'_i-\vecp_i))~w_{u\,\xi_i}\big)
\big(w_{s\,-\xi_k}^*\bm{\sigma} ~w_{s\,\xi_j}\big)
\Big\} + O\Big(\big({p\over m}\big)^2\Big),
\end{eqnarray}
where $w_{u \xi}$ is 2-component spinor of the u-quark, 
and $w^*_{s -\xi}$ is 2-component spinor of the $\sbar$-antiquark.
The SU(3) generator of the color space, $\lambda_i/2$, is to be evaluated 
between the initial and final states of the $i$-th quark. The operator denoted by 
$\lamjkbar/2$ should be evaluated between the $k$-th antiquark and $j$-th quark,
which corresponds to the q$\qbar$ pair annihilating at the vertex creating the gluon.

Using the momentum conservation, $\vecp'_i=\vecp_i+\vecp_j+\vecp_k$, with the notation $\veck = \vecp_j+\vecp_k$,
we obtain the potential as:
\begin{eqnarray}
V_{i;j\overline{k}} &=& \lambda_i\cdot \lamjkbar\;
{\alpha_s\over 4} {\pi\over m_a^2} \Big\{\Big(
{\veck\over 2 m_a}
-{\vecp_i+\vecp'_i +i \bm{\sigma}_i\times\veck \over 2m_i}\Big)\cdot \sigjkbar
\Big\} ~\delta^f_{\overline{k} j},
\end{eqnarray}
where  $\sigjkbar$ is the spin operator  which operates 
between the $k$-th antiquark and $j$-th quark,
and $\delta^f_{\overline{k} j}$ stands for 
that the flavor of the quark of the annihilating 
 pair is equal to that of the antiquark.

Now let us evaluate the matrix element between the q$^3$-pole and the baryon-meson state.
The q$^3$-pole is denoted as $|\Lambda^1(123)\ket$, being assumed as
the flavor-singlet spin-1/2 $(0s)^2(0p)$ state of the three quarks
numbered from 1 to 3.
The baryon in the scattering state is  denoted as $|B(123)\ket$, which is 
antisymmetrized among three quarks
with appropriate quantum numbers. The meson is denoted as $|M(4\overline{5})\ket$,
the q$\qbar$ state with 4th quark and the antiquark numbered as $\overline{5}$.
Then the transfer matrix element can be expressed as
\begin{eqnarray}
\bra \Lambda^1(123)|\sum^4_{i< j}V_{i;j\overline{5}}\,{\cal A}_4|B(123)\,M(4\overline{5})\ket
&=&6 \; \bra \Lambda^1(123)|V_{3;4\overline{5}}{1\over\sqrt{4}}(1-2P_{24}-P_{34})|B(123)\,M(4\overline{5})\ket~,
\end{eqnarray}
where ${\cal A}_4$ is the antisymmetrizing operator over the four quarks, 
and $P_{ij}$ is the exchange operator
between the $i$-th and $j$-th quarks.  
The operator is evaluated in each color, flavor, spin and orbital space.
The direct term vanishes
because we assume that the one-gluon exchange induces the annihilation of the color 
singlet q$\qbar$ pair.
Also, we assume the total momentum of the system
is equal to zero. Since it is enough to calculate 
the operator $V_{3;4\overline{5}}$, only the mixed symmetric (MS) term in the orbital 
space is relevant for the q$^3$ state. So the transfer matrix element becomes
\begin{eqnarray}
\lefteqn{\bra \Lambda^1(123)|V_{3;4\overline{5}} P_{ij}|B(123)\,M(4\overline{5})\ket}&&\nonumber\\
&=&-\sqrt{3}\;V^{tr}_0\sum_\alpha {1\over \sqrt{2}}
\bra (0s)^2(0p)\text{MS};L_z\!=\!1|{\cal O}_\alpha^{orb} |(0s)^5\ket 
\bra \Lambda^1 \text{MS};m_z\!=\!{1\over 2}|{\cal O}_\alpha^{f\sigma} P_{ij}^{f\sigma}| BM\ket 
\bra \lambda_i\cdot \lamjkbar P_{ij}^c \ket
\label{eq:A10}
\end{eqnarray}
with
\begin{eqnarray}
V^{tr}_0 &=& {\pi\alpha_s\over 8 m_u^3}~,~~~
{\cal O}_1^{orb} =
\veck, ~~~
{\cal O}_2^{orb} = 
-(\vecp_3+\vecp'_3), ~~~
{\cal O}_3^{orb} = 
\veck, 
\\
{\cal O}_1^{f\sigma} &=&  \xi_a^3\;\bm{\sigma}_{\overline{5}4}~\delta^f_{\overline{5}4},~~~
{\cal O}_2^{f\sigma} = \xi_a^2\xi_3\; \bm{\sigma}_{\overline{5}4}~\delta^f_{\overline{5}4},~~~
{\cal O}_3^{f\sigma} = \xi_a^2\xi_3\; i \bm{\sigma}_3\times\bm{\sigma}_{\overline{5}4}~\delta^f_{\overline{5}4},~~~
\xi_a = {m_u \over m_a},~~ \text{and} ~~\xi_3 = {m_u \over m_3}~.
\end{eqnarray}
As for the orbital part,
since we assume only from  the $(0s)^5$ component of the baryon-meson wave function 
couples to the q$^3$-pole, we omit the exchange operator from the above equation.
Then,
\begin{eqnarray}
\bra {\cal O}_1^{orb} \ket &=&\bra {\cal O}_3^{orb} \ket= 2A,~~~
\bra {\cal O}_2^{orb} \ket= -4A, ~~~\text{with}~~~
A= -{15^{3/4}\over \sqrt{7 \pi}^3} {\sqrt{3}\over 7}{1\over b^4}.
\end{eqnarray}
The matrix elements  can be calculated straightforwardly.
Color and flavor-spin parts of the matrix elements are shown in 
Table \ref{tbl:V3q5q-c}. 

\begin{table}[tdp]
\caption{Color and flavor-spin parts of the matrix elements in eq.\ (\ref{eq:A10}).}
\begin{center}
{
\begin{tabular}{ccc}
\hline
$\bra(\lambda_3\cdot\lambda_{\overline{5}4})P_{24}\ket$ &
$\bra(\lambda_3\cdot\lambda_{\overline{5}4})P_{34}\ket$ \\[2ex]
$\displaystyle-{8\over 3\sqrt{3}}$ & $\displaystyle{16\over 3\sqrt{3}}$ \\[2ex] \hline
\end{tabular}

\vspace{5mm}

\begin{tabular}{ccccccc}
\hline
&
$\bra {\cal O}_1^{f\sigma} P_{24}\ket$ &
$\bra {\cal O}_1^{f\sigma} P_{34}\ket$ &
$\bra {\cal O}_2^{f\sigma}P_{24}\ket$ &
$\bra {\cal O}_2^{f\sigma}P_{34}\ket$ &
$\bra {\cal O}_3^{f\sigma}P_{24}\ket$ &
$\bra {\cal O}_3^{f\sigma}P_{34}\ket$ 
\\[2ex]
$\bra \Lambda^1_{~\text{MS}}|{\cal O}^{f\sigma}\,|\Sigma\pi\ket$ & 
$\displaystyle {1\over 4}$ & $\displaystyle -{1\over 2}$ &
$\displaystyle {1+2\xi\over 12}$ & $\displaystyle -{1\over 2}$ &
$\displaystyle -{1+2\xi\over 6}$ & 1 \\[3ex] 
$\bra \Lambda^1_{~\text{MS}}|{\cal O}^{f\sigma}\,|{\rm N\overline{K}}\ket$ & 
$\displaystyle-{1\over 2\sqrt{6}}$ & $\displaystyle{1\over \sqrt{6}}$ &
$\displaystyle-{1\over 2\sqrt{6}}$ & $\displaystyle{\xi\over \sqrt{6}}$ &
$\displaystyle {1\over \sqrt{6}}$ & $\displaystyle-{2\xi\over \sqrt{6}}$ 
\\[3ex] 
$\bra \Lambda^1_{~\text{MS}}|{\cal O}^{f\sigma}\,| \Lambda\eta_{ud}\ket$ & 
$\displaystyle  {1\over 12}$ & $\displaystyle  -{1\over 6}$ &
$\displaystyle  {1\over 12}$ & $\displaystyle  -{1\over 6}$ &
$\displaystyle  -{1\over 6}$ & $\displaystyle  {1\over 3}$ 
 \\[2ex]
$\bra \Sigma^8_{~\text{MS}}|{\cal O}^{f\sigma}\,|\Lambda\pi\ket$ & 
$\displaystyle -{1\over 6\sqrt{3}}$ & $\displaystyle {1\over 3\sqrt{3}}$ &
$\displaystyle {1-2\xi\over 6\sqrt{3}}$ & $\displaystyle {1\over 3\sqrt{3}}$ &
$\displaystyle -{1-\xi\over 3\sqrt{3}}$ & $\displaystyle -{2\over 3\sqrt{3}}$ \\[3ex] 
$\bra \Sigma^8_{~\text{MS}}|{\cal O}^{f\sigma}\,|\Sigma\pi\ket$ & 
$\displaystyle -{1\over 9\sqrt{2}}$ & $\displaystyle {\sqrt{2}\over 9}$ &
$\displaystyle {1-2\xi\over 9\sqrt{2}}$ & $\displaystyle {\sqrt{2}\over 9}$ &
$\displaystyle {\sqrt{2}\xi\over 3}$ & $\displaystyle -{2\sqrt{2}\over 9}$ \\[3ex] 
$\bra \Sigma^8_{~\text{MS}}|{\cal O}^{f\sigma}\,|{\rm N\overline{K}}\ket$ & 
$\displaystyle {\sqrt{2}\over 9}$ & $\displaystyle -{2\sqrt{2}\over 9}$ &
$\displaystyle {\sqrt{2}\over 9}$ & $\displaystyle -{2\sqrt{2}\xi\over 9}$ &
$\displaystyle -{1\over 3\sqrt{2}}$ & $\displaystyle {4\sqrt{2}\xi\over 9}$ 
\\[3ex] 
 \hline
\end{tabular}

}\end{center}
\label{tbl:V3q5q-c}
\end{table}%

\section{GCM approach to the baryon meson scattering}
In order to solve the RGM equation, we employ the generator 
coordinate method (GCM).  First we expand the relative wave 
function in terms of locally peaked gaussians centered at 
$\xbld{R}_i$ with the size parameter $B=\sqrt{5/6}b$ as follows: 
\begin{equation}
\chi(\xbld{R}_{BM})=\sum^n_{i=1} C_i 
g(\xbld{R}_{BM}-\xbld{R}_i,B=\sqrt{\frac{5}{6}}b). \label{relwf}
\end{equation}

Because the relative wave function is expanded in terms of the locally 
peaked gaussian, this method can be applied to the bound state 
problem. The modification necessary for
treating the scattering problem will be 
explained later. 

The binding energy $E$ and the expansion coefficients $C_i$ are 
given by the eigenvalues and eigenvectors of 
the following GCM equation:
\begin{equation}
\sum_{j=1}^n H_{ij}C_j
= E\sum_{j=1}^n N_{ij}C_j,  \label{gcm}
\end{equation}
where $n$ is a dimension of the GCM kernels whose matrix 
elements are given by
\begin{equation}
\left\{ \begin{array}{c}
H_{ij} \\ N_{ij}  \end{array}\right \}
=\int \phi^\dagger_{BM}(\xbld{R}_i)
\left\{ \begin{array}{c}
H \\ 1  \end{array} \right\}
{\cal A}
\phi_{BM}(\xbld{R}_j) {\displaystyle \prod^5_{k=1}d\xbld{r}_k} 
\label{gcmk}.
\end{equation}
Here the $\phi_{BM}(\xbld{R}_i)$ is the five quark (4 quarks and 1 antiquark) wave function 
whose orbital part $\varphi_{BM}(\xbld{R}_i)$ is given by 
the following product:
\[
\varphi_{BM}(\xbld{R}_i)=\varphi_B(\xbld{\xi}_B)\varphi_M(\xbld{\xi}_M)
 g(\xbld{R}_{BM}-\xbld{R}_i,B=\sqrt{\frac{5}{6}}b) g(\xbld{R}_G,B=\sqrt{\frac{1}{5}}b).
\]
When we rewrite the integrals over the internal and relative coordinates 
in terms of single quark coordinates in eq.(\ref{gcmk}), we have employed the 
following equation for the center of mass coordinate $\xbld{R}_G$:
\[
\int g(\xbld{R}_G,B=\sqrt{\frac{1}{5}}b) g(\xbld{R}_G,B=\sqrt{\frac{1}{5}}b) 
d \xbld{R}_G=1.
\]
Employing the gaussian form for the internal wave function,  
the orbital part $\varphi_{BM}(\xbld{R}_i)$ is written as a 
product of single quark wave functions with size 
parameter $b$, centered at $2\xbld{R}_i/5$ and $-3\xbld{R}_i/5$:
\begin{equation}
\varphi_{BM}(\xbld{R}_i)
= \prod^3_{k=1} g(\xbld{r}_k - \frac{2\xbld{R}_i}{5},b) 
\prod^5_{k=4} g(\xbld{r}_k + \frac{3\xbld{R}_i}{5},b). \label{pwf}
\end{equation}
Note that it is quite easy to perform the antisymmetrization 
of the four quarks when the wave
function is given in terms of single-quark coordinates as 
in eq.(\ref{pwf}).

Now let us discuss the renormalization of the wave function.  
Employing the following equation,
\[
H \chi = E N \chi
\rightarrow 
\frac{1}{\sqrt{N}}H\frac{1}{\sqrt{N}}\sqrt{N}\chi=E\sqrt{N}\chi,
\]
we call $\sqrt{N}\chi$ the renormalized wave function
which satisfies the Schr\"{o}dinger equation. 
Using the coefficients $C_i$ given by solving eq.(\ref{gcm}) 
which are normalized as
\[
\sum_{ij}^{n} C_i N_{ij}C_j=1,
\]
the wave function $\sqrt{N}\chi$ is properly orthonormalized.

The modification of the relative kinetic energy for the baryon-meson 
scattering problem explained in section III is rewritten in terms of 
the GCM kernel as follows. We take out the direct term of the kinetic energy operator 
$K_{Dij}$ in the GCM kernel $H_{ij}$.
By subtracting the internal kinetic energy with the direct norm kernel $N_{Dij}$, 
the GCM kernel of the relative kinetic energy is given by 
\[
K_{R ij}=K_{Dij} - 3K_0 \times N_{Dij},  \hspace{10pt}\mbox{where} 
\hspace{10pt}K_0=\frac{3}{4m_q b^2} .
\]
Then the relative kinetic energy $K_{R ij}$ is modified in the following way.
\[
K_{R ij} \rightarrow K_{R ij} \times \frac{6 m_q}{5} \frac{1}{\mu},
\]
where $\mu$ is the reduced mass for the baryon-meson system 
calculated by using the masses given in Table II. 
The GCM kernel of the coupling of the baryon-meson state with the 3q state is taken as
\[
\bra 3q | V_{3q-5q} | BM, R_j\ket=\bra 0s^20p |V_{3q-5q} | 0s^5\ket N_{ij}
\hspace{10pt} {\rm with} \hspace{5pt} R_i=0 .
\]

Now we will explain how to modify the GCM 
in order to treat the scattering 
problem. Because practical calculations of the bound state 
and scattering problem are done in terms of partial waves, 
we first explain the partial wave expansion.
The relative wave function in eq.(\ref{relwf}) is expanded 
in terms of locally peaked wave functions with a definite angular 
momentum $lm$:
\begin{equation}
\chi(\xbld{R}_{BM})=\sum^n_{i=1} C_i 
\chi_i^{(l)}(R_{BM}) Y_{lm}(\hat{\xbld{R}}_{BM}) \label{relpwf}.
\end{equation}
Here, $\chi_i^{(l)}(R)$ is given by an expansion of 
locally peaked gaussian wave functions:
\[
g(\xbld{R}-\xbld{R}_i,B)=
(\sqrt{\pi} B)^{-3/2} \exp(-\frac{(\xbld{R}-\xbld{R}_i)^2}{2B^2})
=\sum_{lm} \chi_i^{(l)}(R)Y_{lm}(\hat{\xbld{R}})
Y_{lm}^{\ast}(\hat{\xbld{R}}_i).
\]

The explicit form of $\chi$ is given by
\[
\chi_i^{(l)}(R)=4 \pi (\sqrt{\pi} B)^{-3/2}\exp (-\frac{R^2+R_i^2}{2B^2})
i_l(\frac{RR_i}{B^2}),
\]
where $i_l$ is the modified spherical Bessel function.  
When we treat the scattering problem, we use the following modified 
wave functions for the expansion
\begin{eqnarray*}
\tilde \chi_i^{(l)}(R)&=& \alpha_i \chi_i^{(l)}(R),\hspace{20pt} (R < R_c) \\\tilde \chi_i^{(l)}(R)&=& h_l^{(2)}(kR) +s_i h_l^{(1)}(kR), 
\hspace{20pt} (R \geq R_c).
\end{eqnarray*}

Here, $k$ is a wave number and $h_l^{(1)}$ and $h_l^{(2)}$ are spherical 
Hankel functions. The coefficients $\alpha_i$ and $s_i$ are determined 
by a continuity condition of  $ \tilde \chi$ at $R=R_c$.
The relative wave function is expanded in terms of $\tilde \chi$ as

\[
\chi^{(l)} (R_{BM})=\sum_{i=1}^n C_i \tilde \chi_i^{(l)}(R_{BM})
\hspace{10pt} {\rm with} \hspace{10pt} \sum_{i=1}^n C_i=1.
\]
Then the $S$-matrix is given in terms of the coefficients $C_i$ as:
\[
S=\sum_{i=1}^n C_is_i.
\]
The details how to determine the expansion coefficients can be found in Oka and Yazaki
\cite{okya}.

\end{document}